\documentclass[reprint, 10pt,a4paper]{revtex4-1}
\usepackage[utf8]{inputenc}
\usepackage{amsmath}
\usepackage{amsfonts}
\usepackage{amssymb}
\usepackage{mathrsfs}
\usepackage{subfigure}
\usepackage{hyperref}

\usepackage{empheq}

\newcommand{\bv}{\mathbf{v}}

\newcommand{\bB}{\mathbf{B}}

\begin{document}

\begin{abstract}
In a variety of magnetized plasma geometries, it has long been known that highly charged impurities tend to accumulate in regions of higher density. Here, we examine how this ``collisional pinch" changes in the presence of additional forces, such as might be found in systems with gravity, fast rotation, or non-negligible space charge. In the case of a rotating, cylindrical plasma, we describe a regime in which the radially outermost ion species is intermediate in both mass and charge. We discuss implications for fusion devices and plasma mass filters. 
\end{abstract}

\title{Strategies for Advantageous Differential Transport of Ions in Magnetic Fusion Devices}
\date{\today}
\author{E. J. Kolmes}
\email{ekolmes@princeton.edu}
\author{I. E. Ochs}
\author{N. J. Fisch}
\affiliation{Department of Astrophysical Sciences, Princeton University, Princeton, New Jersey, USA}
\affiliation{Princeton Plasma Physics Laboratory, Princeton, New Jersey, USA}
\maketitle

\section{Introduction}

The diffusive transport of plasma across a magnetic field is a subject of longstanding importance throughout plasma physics. From tokamaks \cite{Hirshman1981} and stellarators \cite{Braun2010} to magnetic mirrors \cite{Post1987, Gueroult2012} and low-temperature plasmas \cite{Smirnov2004, Ellison2012, Curreli2014}, problems related to cross-field transport are prominent both for their intrinsic scientific significance and for their practical implications in the design and analysis of plasma-based technology. The transport properties of high-$Z_I$ impurities is of particular interest in fusion plasmas, where heavier elements from the wall can reduce the performance of a fusion device. They are also very important in plasma mass filters, which are designed to sort the material in a plasma based on mass \cite{Bonnevier1966, Lehnert1971, Hellsten1977, Krishnan1983, Geva1984, Dolgolenko2017, Ochs2017iii}. 

Early in the development of plasma transport theory, it was predicted that high-$Z_I$ impurities in a predominantly low-$Z_b$ magnetized background plasma would demonstrate a dramatic pinch effect \cite{Spitzer1952, Taylor1961ii, Braginskii1965}. In particular, in steady state, the impurity density $n_I$ and the background ion density $n_{b}$ in an isothermal plasma satisfy 
\begin{gather}
\frac{(n_{b})^{Z_I/Z_{b}}}{n_I} = \text{const.} \label{eqn:pinch}
\end{gather}
If impurities are introduced in small quantities at the low-density edge of a magnetically confined plasma, Eq.~(\ref{eqn:pinch}) implies that they will be strongly concentrated in the high-density core of the plasma \cite{Taylor1961ii}. This same result has been found in neoclassical transport in a wide range of parameter regimes \cite{Connor1973, Rutherford1974, Hinton1974}. 

There have been experimental indications of a pinch of impurities in plasma devices \cite{Fussman1991, Dux2004}. However, the impurity pinch may be mitigated in some practical contexts. For instance, although neoclassical corrections can increase the speed of the pinch, analysis and experiments have shown that the effect might be reduced by temperature gradients in some regimes \cite{Rutherford1974, Wade2000, Dux2004}. Of course, the presence of turbulent transport can change all of these results \cite{Ida2009, Loarte2015}. 

However, Eq.~(\ref{eqn:pinch}) is no longer accurate in the presence of an external potential. 
Certain corrections associated with centrifugal or electrostatic forces acting on a magnetically confined plasma have been studied already, both experimentally and with analytical and computational models. 
This work has mostly been in the context of tokamak physics \cite{Burrell1981, Romanelli1998, Wade2000, Taylor2005, Gourdain2006, Angioni2014, Casson2015} and plasma mass filters \cite{Bonnevier1966, Lehnert1971, Hellsten1977, Krishnan1983, Geva1984, Fetterman2011b, Gueroult2014, Rax2016, Dolgolenko2017}. Here we derive a more general form of Eq.~(\ref{eqn:pinch}) that can account for the presence of an arbitrary external potential. The case of a centrifugal potential is equivalent to an expression used in the context of plasma centrifuges \cite{Krishnan1983, Geva1984, Dolgolenko2017}.

This generalized formulation makes it possible to intuitively describe the differential transport of various ion species in a variety of different systems. 
Beyond being academically interesting, this description serves a practical purpose, revealing regimes in which desirable and undesirable impurities can be differentially transported with greater freedom. 
For instance, in p-$^{11}$B fusion, boron ions and protons fuse to form helium.
In a fully ionized system, both boron and helium ash are more massive and highly charged than the protons. 
However, boron (as a fuel ion) is desirable to concentrate in the core of a fusion device, whereas the accumulation of helium ash would reduce fusion performance. 
Eq.~(\ref{eqn:pinch}) does not offer any way of choosing a proton profile that would draw in boron while pushing out helium. 
The generalized formula does describe a window in parameter space in which such an outcome should be possible. 
Thus understanding the detailed transport behavior of different species in a plasma could be important for increasing fusion efficiency.

The paper is organized as follows.
We begin with a brief derivation of Eq.~(\ref{eqn:pinch}), generalized to include external potentials, and show how the collisional pinch and Gibbs distribution emerge naturally as limits of the resulting impurity distribution.
We then discuss some of the implications of our results in gravitational and centrifugal potentials. 
For the centrifugal potential, we discuss strategies for flushing impurities and fusion products in a couple of different scenarios. 

\section{Derivation of Generalized Pinch} \label{sec:fluid}

Cross-field collisional transport is driven primarily by interactions between unlike particles \cite{Longmire1956}. Consider a plasma with two ion species, indexed by subscripts $I$ and $b$, in a magnetic field with no externally imposed forces other than $\mathbf{j} \times \mathbf{B}$ forces. If this system contains pressure gradients, some cross-field motion will occur via diamagnetic drifts. This motion will be perpendicular both to $\nabla p$ and to $\mathbf{B}$. In a cylindrical geometry with an axial magnetic field and $p = p(r)$, it will result in azimuthal flows. 

When there are diamagnetic drifts in a collisional magnetized plasma, there is an additional cross-field transport mechanism that arises from the relative diamagnetic motion of the particle species. This relative motion causes a frictional force, which in turn drives an $\mathbf{F} \times \mathbf{B}$ drift parallel to the pressure gradients. Eq.~(\ref{eqn:pinch}) can be derived from the condition that the frictional forces between these two species vanish (i.e., that there is no relative velocity). This derivation has been done previously in both the fluid picture \cite{HelanderSigmar} and the single-particle picture \cite{Ochs2017}. Here we will briefly replicate and extend the argument in the fluid picture, allowing for an additional species-dependent external potential $\Phi_s$ with a gradient parallel to that of the pressure. 

The force felt by a particle of species $I$ as a result of friction with another species $b$ can be modeled as $-m_I \nu_{Ib} (\bv_I - \bv_b)$. The force on species $b$ is the same up to exchange of indices, where the conservation of momentum requires that the collision frequencies satisfy $n_I m_I \nu_{Ib} = n_b m_b \nu_{bI}$. 
Consider a system without temperature gradients or shear strain forces, but with density gradients perpendicular to $\bB$. Assume the different species all have the same temperature. The fluid momentum equation is
\begin{align}
m_I \frac{d \bv_I}{dt} = q_I \bv_I \times \bB - \frac{T \nabla n_I}{n_I} - \nabla \Phi_I \nonumber \\
+ \sum_{s} m_I \nu_{I s} (\bv_{s} - \bv_I). 
\end{align}
From this, and neglecting the inertial terms, the flux of species $I$ in the direction parallel to the pressure gradients due to collisions with species $b$ is 
\begin{align}
\Gamma_I &= \frac{1}{2} \rho_I^2 n_I \nu_{Ib} \bigg\{ \bigg[ \frac{\nabla n_I}{n_I} - \frac{Z_I}{Z_{bg}} \frac{\nabla n_b}{n_b} \nonumber \\
&\hspace{70 pt}+ \frac{\nabla \Phi_I}{T} - \frac{Z_I}{Z_b} \frac{\nabla \Phi_b}{T} \bigg] \times \hat{b} \bigg\} \times \hat{b}. 
\label{eqn:flatTempClassicalFlux}
\end{align}
Here $\rho_I$ is the Larmor radius for species $I$ and $\hat{b} \doteq \mathbf{B} / B$. 
When there is an external potential, the relative motion between the species depends on both diamagnetic and $-\nabla \Phi \times \mathbf{B}$ drifts. 

When $\Gamma_I$ vanishes, Eq.~(\ref{eqn:flatTempClassicalFlux}) is equivalent to the condition that 
\begin{equation}
\left( n_I e^{\Phi_{I} / T} \right) \propto \left( n_{b} e^{\Phi_{b} / T} \right)^{Z_I / Z_{b}}. \label{eqn:phiPinch}
\end{equation}
Eq.~(\ref{eqn:phiPinch}) is the generalization of Eq.~(\ref{eqn:pinch}) in the presence of an external potential. When $\nabla \Phi_b = \nabla \Phi_I = \mathbf{0}$, it reduces to the original pinch. If the background profile is completely supported by the potential $\Phi$ (that is, if $n_b$ assumes a Gibbs distribution) then the spatial dependence on the right-hand side cancels, so the impurity species must also be Gibbs-distributed. For cases between the Gibbs distribution and the original pinch, Eq.~(\ref{eqn:phiPinch}) describes a whole family of solutions that fall between the Gibbs distribution and the classic pinch. If only part of the pressure profile is supported by the potential $\Phi$, Eq.~(\ref{eqn:phiPinch}) implies that the pinch acts only on the part of the profile not supported by $\Phi$. That is, if $n_s = \tilde{n}_s e^{-\Phi_s / T}$, then $\tilde{n}_I \propto \tilde{n}_b^{Z_I/Z_b}$. Interestingly, if $\Phi_I / \Phi_b = Z_I / Z_b$, then Eq.~(\ref{eqn:phiPinch}) reduces to Eq.~(\ref{eqn:pinch}); an electrostatic potential and the collisional pinch both tend to concentrate highly charged particles, but they do so to the same extent and their effects do not stack. 

Nothing in this derivation has assumed anything in particular about the relative masses, densities, or charges of the species $I$ and $b$, so long as all species are magnetized. However, the flux of species $I$ due to interactions with species $b$ scales with $\nu_{Ib}$. If Eq.~(\ref{eqn:pinch}) is not satisfied for a pair of species $I$ and $b$, the size of the resulting flux will depend on their collision frequency. In other words, when $\nu_{Ib}$ is larger, the system will tolerate smaller deviations from Eq.~(\ref{eqn:pinch}) in order to be in steady state on any given timescale.

For this reason, it often makes sense to treat Eq.~(\ref{eqn:pinch}) as a requirement for ion density profiles with respect to one another rather than for ion profiles with respect to the electron profile. 
Of course, predictions of very large charge separations may not be realistic, especially once the resulting electric fields would grow large enough to interfere with the assumed gyromotion of the particles. 

It is important to note that Eq.~(\ref{eqn:phiPinch}) is not, on its own, a prescription for how $n_I$ reacts to the introduction of an external potential. It describes a relationship between $n_I$ and $n_b$; finding $n_I$ requires specifying $n_b$. As the strength of the potential $\Phi$ varies, Eq.~(\ref{eqn:phiPinch}) predicts very different behavior of the impurity profile depending on how the background ion profile changes. 

\section{Distribution Limits in a Simple Linear Potential}\label{sec:gravity}

Consider a multiple-species plasma with some species-dependent potential $\Phi_s$ and a constant magnetic field $\mathbf{B} = B \hat{z}$. If every species satisfies Eq.~(\ref{eqn:phiPinch}) with respect to a fixed reference species, then they automatically satisfy Eq.~(\ref{eqn:phiPinch}) with respect to one another. Suppose some reference ion species has mass $m_r$ and charge $Z_r e$. Then define a ``steepness parameter" $\alpha_r$ for which 
\begin{gather}
n_r \propto \exp \bigg[ - \frac{\alpha_r \Phi_r}{T} \bigg]. \label{eqn:referenceSpeciesDefinition}
\end{gather}
In fact, $\alpha_r$ could be a function rather than a constant, and the results in this section would continue to hold. The steepness parameter $\alpha_r$ measures how steep the reference distribution is relative to its thermodynamic equilibrium distribution. 

Eq.~(\ref{eqn:phiPinch}) can be rewritten in terms of a set of steepness parameters as 
\begin{gather}
n_s \propto \exp \bigg[ - \frac{\alpha_s \Phi_r}{T} \bigg] \\
\alpha_s = \frac{Z_s}{Z_r} \big( \alpha_r - 1 \big) + \frac{\Phi_s}{\Phi_r}. \label{eqn:generalSteepness}
\end{gather}
Consider, for example, a plasma in a linear gravitational potential $\Phi_s = m_s g x$. For this choice of potential, Eq.~(\ref{eqn:phiPinch}) becomes 
\begin{gather}
n_s = (\text{const}) \, n_r^{Z_s/Z_r} \exp \bigg[ Z_s \bigg( \frac{m_r}{Z_r} - \frac{m_s}{Z_s} \bigg) \frac{g x}{T} \bigg] \label{eqn:gravityPinch}
\end{gather}
and Eq.~(\ref{eqn:generalSteepness}) becomes 
\begin{gather}
\alpha_s = \frac{Z_s}{Z_r} \big( \alpha_r - 1 \big) + \frac{m_s}{m_r}. \label{eqn:massDependentSteepness}
\end{gather}
It is immediately apparent that both the charge ratio $Z_I / Z_b$ and mass ratio $m_I / m_b$ now play a role in determining the impurity distribution. 
Define the thermodynamic scale height $\lambda_s$ by 
\begin{equation}
	\lambda_s \doteq \frac{T}{m_s g}. 
\end{equation}
For the gravitational potential, any constant $\alpha_s$ has an intuitive interpretation: it is the inverse scale height of the distribution, normalized by $\lambda_r$. 

Take $\alpha_r$ as a free parameter that is imposed; in effect, it represents how far a reference profile is from thermodynamic equilibrium, with $\alpha_r = 1$ corresponding to thermodynamic equilibrium.
A specific choice of $\alpha_r$ then imposes a different $\alpha_s$ on each other plasma species via Eq.~(\ref{eqn:massDependentSteepness}).

Eq.~(\ref{eqn:massDependentSteepness}) shows that the balance between the charge-dependent and mass-dependent parts of $\alpha_s$ depends critically on the magnitude of $\alpha_r$. 
In the potential-free limit $\alpha_r \rightarrow \infty$, Eq. (\ref{eqn:massDependentSteepness}) becomes
\begin{gather}
\alpha_s = \frac{Z_s}{Z_r} \alpha_r,
\end{gather}
which is simply a reformulation of the collisional pinch Eq.~(\ref{eqn:pinch}). 
If instead $\alpha_r =1$, i.e. the reference species has reached thermodynamic equilibrium, then 
\begin{gather}
\alpha_s = \frac{m_s}{m_r},
\end{gather}
so that 
\begin{equation}
n_s(x)  = n_{s0} e^{-x m_s / m_r \lambda_r} =  n_{s0} e^{-x /\lambda_s},
\end{equation}
where $\lambda_s=T / m_s g$ is the thermodynamic scale height for the impurity in the gravitational field. 
In other words, as the background approaches thermodynamic equilibrium (for instance, due to collisions with electrons), the impurity naturally approaches thermodynamic equilibrium as well.

\section{Centrifuging Ion Species}\label{sec:centrifuge}

\begin{figure*}
	\includegraphics[width=.9\linewidth]{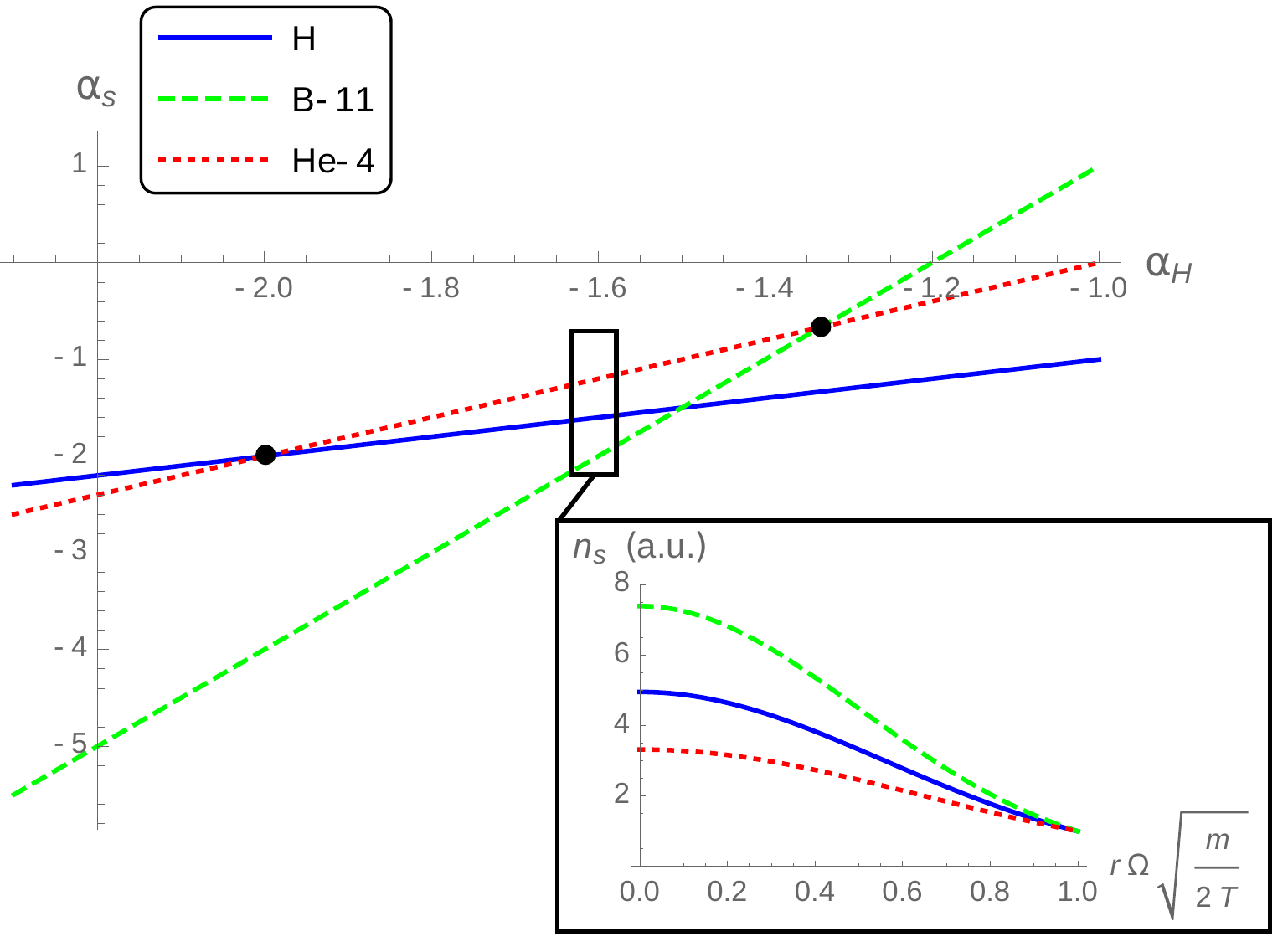}
	\caption{The steepness parameters $\alpha_s$ of three particle species as a function of the hydrogen steepness parameter $\alpha_H$ in a centrifugal potential. A more negative $\alpha_s$ corresponds to a profile that is more peaked at small $r$. The region in which the alpha particles are less concentrated in the core than the proton and $^{11}$B populations (despite the fact that the alpha particles are intermediate in charge and mass) is $-2 < \alpha_H < -4/3$. This region is marked by dots. The inset figure shows spatial profiles corresponding to a choice of $\alpha_H = -1.6$, which falls in this region; it can be seen that the $^{11}$B is twice as peaked relative to its edge density as the $^{4}$He. \label{fig:steepness}}
\end{figure*}

In a rotating system with several ion species, the balance between charge-dependent and mass-dependent effects can lead to useful outcomes. 
Consider an isothermal cylindrical system undergoing solid-body rotation, so that there is an effective centrifugal potential $\Phi(r) = - m \Omega^2 r^2 / 2$. In terms of an arbitrary reference species $r$, Eq.~(\ref{eqn:phiPinch}) becomes 
\begin{gather}
n_s = (\text{const}) \, n_r^{Z_s/Z_r} \exp \bigg[ Z_s \bigg( \frac{m_s}{Z_s} - \frac{m_r}{Z_r} \bigg) \frac{\Omega^2 r^2}{2 T} \bigg]. \label{eqn:centrifugalPinch}
\end{gather}
That is, 
\begin{align}
\bigg( n_s \exp \bigg[ &- \frac{m_s \Omega^2 r^2}{2 T} \bigg] \bigg)^{1/Z_s} \nonumber \\
&= (\text{const}) \bigg( n_r \exp \bigg[ - \frac{m_r \Omega^2 r^2}{2 T} \bigg] \bigg)^{1/Z_r}. \label{eqn:centrifugalPinchResorted}
\end{align}
This is equivalent to a previously derived result \cite{Krishnan1983, Geva1984}. Given a fixed $n_r$, whether the distribution $n_I$ will be steeper or less steep than it would have been without rotation depends on the relative charge-to-mass ratios of species $s$ and $I$, as was true in the gravitational case. 

For a centrifugal potential,  Eq.~(\ref{eqn:referenceSpeciesDefinition}) is 
\begin{gather}
n_r \propto \exp \bigg[ \frac{\alpha_r m_r \Omega^2 r^2}{2 T} \bigg]. \label{eqn:alphaRDefinition}
\end{gather} 
Due to centrifugal forces, this would mean that for $\alpha_r =1$  the reference ions are distributed according to their thermal equilibrium, namely, flung towards high radius, with minimum density at $r=0$.  If the reference ions are so distributed, then by Eq.~(\ref{eqn:centrifugalPinchResorted}), it is immediately evident that so are all ions.  For $\alpha_r >1$, the reference ions are more concentrated yet at large radius, which means that, relatively speaking, all ions are more concentrated at large radii compared to their thermal equilibrium distribution.  For $\alpha_r < 1$, ions are less steeply distributed.  However, if $\alpha_r < 0$, the ions are inverted with respect to their distribution in thermal equilibrium; in this case, the distribution is peaked at the center rather than at the periphery. This is the case of general interest in fusion devices, where fusion occurs preferentially in a central hot and dense core of plasma. 

Note that not all the $\alpha_i$ need have the same sign, or if they do have the same sign, their relative ordering can be a function of $\alpha_r$.  It is of interest to arrange for fuel ions to be concentrated preferentially in the interior near $r=0$, while fusion byproducts or other contaminants are comparatively less concentrated in the interior. For the centrifugal potential, $\Phi_s / \Phi_r = m_s / m_r$, so Eq.~(\ref{eqn:massDependentSteepness}) is the governing equation for the steepness parameters. 

Since the reference species $r$ is arbitrary, we can take it to be protons (whether or not protons are actually present is immaterial; we can still treat $\alpha_H$ as a free parameter of the system). With that choice, Eq.~(\ref{eqn:massDependentSteepness}) becomes
\begin{gather}
\alpha_s = Z_s (\alpha_H - 1 \big) + \frac{m_s}{m_p}. 
\end{gather}
As was the case with the gravitational potential, there are two limits: one in which the charge-dependent term dominates and the result approaches Eq.~(\ref{eqn:pinch}), and another in which the mass-dependent term dominates and the result approaches a Gibbs distribution. 

When $\alpha_r < 1$, it can happen that the species that is radially the furthest out has neither the highest $Z_s$ nor the highest $m_s$. For instance, consider a fully ionized plasma of protons, boron-11, helium-4, and tungsten-184. This might model a p-$^{11}$B fusion plasma with thermalized fusion products and some tungsten impurities. In a non-rotating system, the outermost species would always be the p or the $^{184}$W, since these have the smallest and largest charges, respectively. In a rotating plasma in thermal equilibrium (i.e. in a Gibbs distribution), the outermost species would always be the $^{184}$W, because of its large mass. But between these two limits, it is possible to find cases in which any of these species are the furthest out, including the $^{4}$He. This result could not be achieved using just the physics of the conventional pinch or the physics of a conventional centrifuge. Essentially, it is possible when the classical pinch and the centrifugal force pull in opposite directions. 

Fig.~\ref{fig:steepness} shows the relative steepness of the hydrogen, helium, and boron for a variety of choices of $\alpha_H$. In fusion devices, it is typically advantageous to limit the buildup of fusion products and impurities like tungsten relative to the fuel ions. 
There is a region in Fig.~\ref{fig:steepness} in which the fuel ions and thermalized $^{4}$He particles are all peaked toward small $r$ but where the $^{4}$He are less concentrated in the core than the fuel ions. This happens when 
\begin{gather}
-2 < \alpha_H < - \frac{4}{3}. \label{eqn:alphaRegime}
\end{gather}

There is something special about the case of $\alpha$ ash in a p-$^{11}$B plasma: the fusion product has a charge that is between the charges of the two fuel ions. Consider a rotating system with ion species $a$, $b$, and $c$, in which $Z_a < Z_b < Z_c$. Then there exists a choice of $\alpha_a$ for which $\alpha_b > \alpha_a, \alpha_c$ if and only if
\begin{gather}
\frac{m_c - m_b}{Z_c-Z_b} < \frac{m_b - m_a}{Z_b - Z_a} . 
\end{gather}
For any reference species $r$, the width of this region will be 
\begin{gather}
\Delta \alpha_r = \frac{Z_r}{m_r} \bigg[ \frac{m_b - m_a}{Z_b - Z_a} - \frac{m_c - m_b}{Z_c - Z_b} \bigg]. 
\end{gather}

Very similar analyses can be done for a variety of choices of fuel and impurity ions. 
For instance, for a p-$^{11}$B plasma, there is also a somewhat smaller region (not marked in the figure) in which the fusion products and tungsten impurities are both further out than the fuel ions. This happens when
\begin{gather}
	- \frac{110}{73} < \alpha_H < - \frac{4}{3}. \label{eqn:bothRegime}
\end{gather}
The large $Z$ of the impurities makes the transition between inwardly- and outwardly-peaked $^{184}$W quite abrupt. Of course, this is all for fully ionized tungsten; partially ionized tungsten would have a larger window due to its lower charge state. 

In a bulk deuterium-tritium plasma, thermalized $\alpha$ particles will be radially further out than the fuel ions if any of the following (equivalent) conditions are met, continuing to take protons as the reference species: 
\begin{gather}
\alpha_\text{D} > 1 \\
\alpha_\text{T} > 2 \\
\alpha_{^{4}\text{He}} > 2. 
\end{gather}
Rather than a window, the requirement is a simple inequality. The favorable ion distributions happen when both fuel and fusion product profiles are peaked at the radial edges of the device, though they do not have to be as peaked as they would be in thermal equilibrium. 

In a rotating plasma composed mostly of deuterium and $^{3}$He, the condition for thermalized $\alpha$ particles to be the furthest out can be written in any of the following ways: 
\begin{gather}
\alpha_D > 0 \\
\alpha_{^{3}\text{He}} > -1 \\
\alpha_{^{4}\text{He}} > 0. 
\end{gather}
This is qualitatively similar to the condition for a D-T plasma. However, D-$^{3}$He reactions also produce protons, and there is no choice of steepness parameters for which thermalized protons will be radially further out than deuterium. 

For most realistic scenarios, these regions in $\alpha_r$ space don't require very steep profiles. If a plasma has much less energy in its rotational motion than in its thermal motion, and if the steepness parameters are $\sim \pm 1$, Eq.~(\ref{eqn:alphaRDefinition}) implies that $n$ will not be very steep for any of the species. For instance, for a 1~keV cylindrical plasma with $\Omega = 100$~kHz, the proton profile corresponding to $\alpha_r = -3/2$ would be 
\begin{gather}
n_p \approx n_{p0} e^{- (r / 3.6 \text{ m})^2}. 
\end{gather}
In other words, even a very high rotation rate would not require unrealistically steep profiles in order to fall in the regimes described by, e.g., Eq.~(\ref{eqn:alphaRegime}) and Eq.~(\ref{eqn:bothRegime}). 

Of course, this kind of calculation could also be useful for plasma mass filter applications. For instance, in order to drive a heavy impurity with $m_I > m_b$ and $Z_I > Z_b$ radially outwards, it is better to have a background profile that is flat than one that is peaked at the core, since this prevents the pinch effect from competing with the centrifugal effects. This kind of question has been studied experimentally \cite{Skibenko2009}; the work by Skibenko \textit{et al.} included consideration of the role of $n_s(\mathbf{r})$ in these problems, though not the role of $\Phi$. 

It is important to note that the analysis in this section has not included the effects of temperature gradients or net particle fluxes. Either of these could be important for the behavior of a real device. Nonetheless, the simple calculation presented here raises some interesting practical possibilities. 

\section{Discussion}

In the presence of an external potential, the collisional impurity pinch contains two parts. One part depends on the impurities' charge and on how close the background ions are to thermodynamic equilibrium. The other part depends only on the external potential acting on the impurities. It is possible to affect the balance between these terms by changing the external potential or by changing the background ion density profile. The ability to predict and control the behavior of the pinch could be of use in any application where impurity concentrations matter. 

These results relate the different density profiles to one another, but they do not fully specify the different profiles. To do that would require two things. First, there would have to be a way of specifying the normalization of each profile; this might be done with a boundary condition or with a condition on the total number of particles. Second, one profile (or linear combination of profiles) must be determined independently. Control over this profile is a practical problem which we do not address here. It might be done with neutral particle sources, like pellet injection, or through waves, as in alpha channeling \cite{Fisch1992}. Both of these processes will be balanced by ambipolar particle diffusion out of the device. We also do not address the problem of how to set up a potential $\Phi$. A centrifugal potential, for instance, might be set up by generating perpendicular $\mathbf{E}$ and $\mathbf{B}$ fields. Rotation profiles can also be manipulated with compressional techniques \cite{Geyko2013, Geyko2017}. 

Of course, the calculations presented here do not present a comprehensive picture of all ways to modify the impurity pinch and their implications. This paper does not discuss the implications of impurities that are hotter or colder than the background plasma, nor have we dealt with spatial temperature gradients. Temperature gradients are known to affect the impurity pinch in different ways, depending on the details of the system \cite{Rutherford1974, Hinton1974, Hinton1976, HelanderSigmar}. This is an area of active research; for instance, there has been recent progress on the possibility of using temperature gradients to mitigate the impurity pinch in stellarators \cite{Helander2017, Newton2017}. 

These calculations are also far from being a comprehensive treatment of the effects of rotation and of electrostatic potentials in a practical device. For instance, the tendency of rotational effects to cause uneven distributions across flux surfaces in a toroidal geometry can be important in some systems \cite{Romanelli1998, Angioni2014}. We've also neglected the effects of viscosity; this is safe for systems sufficiently close to solid-body rotation, but not for more general rotation profiles. 

This study was motivated in part by a larger investigation of the Wave-Driven Rotating Torus (WDRT) plasma confinement concept. In a WDRT, minor-radial electric fields and toroidal magnetic fields provide the rotational transform by setting up $\mathbf{E} \times \mathbf{B}$ rotation \cite{Rax2017, Ochs2017ii}. The transport of minority ions in such a device is complicated but important, since one way of maintaining the large voltage gradients would be to preferentially remove helium ash from the device (by $\alpha$ channeling or otherwise). The results discussed here suggest that electric fields themselves won't affect the impurity accumulation in a WDRT, and that the rotation induced by the crossed fields will change the impurity transport in a way that might be advantageous. 

\acknowledgments
We would like to acknowledge S. Davidovits, A. S. Glasser, M. Mlodik, R. Gueroult, and J.-M. Rax for helpful conversations. 
This work was supported by DOE Grants DE-SC0016072, DE-FG02-97ER25308, and DE-AC02-09CH1-1466. 

\bibliographystyle{apsrev4-1} 
\bibliography{../../../Master}

\end{document}